\documentclass[10pt,%
superscriptaddress,
 amsmath,amssymb,longbibliography,
 aps,prfluids,
]{revtex4-2}

\usepackage{graphicx}
\usepackage{dcolumn}
\usepackage{bm}
\usepackage{siunitx}
\usepackage{multirow} 
\usepackage{verbatim}
\usepackage{setspace}

\begin{document}

\preprint{APS/123-QED}

\title{Tuning laser-induced optical breakdown and cavitation through the ionic environment in aqueous media} 




\author{Junhao Cai}
\affiliation{School of Ocean and Civil Engineering, Shanghai Jiao Tong University, Shanghai 200240, China}

\author{Yuhan Li}

\author{Yunqiao Liu}

\author{Benlong Wang}

\author{Mingbo Li}
\email{mingboli@sjtu.edu.cn}
\affiliation{School of Ocean and Civil Engineering, Shanghai Jiao Tong University, Shanghai 200240, China}
\affiliation{Key Laboratory of Hydrodynamics (Ministry of Education), Shanghai Jiao Tong University, Shanghai 200240, China}


\date{\today}

\begin{abstract}
Laser-induced cavitation in liquids originates from optical breakdown processes that depend sensitively on both laser–plasma dynamics and the chemical microenvironment of the solvent.
Herein, we experimentally decouple the effects of ionic strength and ion specificity on cavitation inception in aqueous electrolytes spanning neutral, acidic, and alkaline regimes.
Using focused nanosecond laser pulses, we show that increasing ionic strength universally lowers the cavitation threshold by enhancing charge screening and seed-electron availability.
However, under constant ionic strength, strongly asymmetric behavior emerges: acidic (hydrogen chloride, HCl) solutions inhibit cavitation, whereas alkaline (sodium hydroxide, NaOH) solutions enhance it.
This asymmetry arises from hydrated-electron kinetics that depends on the ion specificity. 
In acidic solutions, hydronium ions act as diffusion-limited scavengers of hydrated electrons, quenching their lifetime and inhibiting avalanche ionization.
In contrast, hydroxide ions reduce hydronium availability and extend electron survival, promoting more efficient plasma formation.
Despite large discrepancies in breakdown thresholds, electrical conductivity remains nearly constant, demonstrating that microscopic electron chemistry, rather than bulk charge transport, governs cavitation onset.
These results establish a mechanism connection between electrolyte chemistry and optical breakdown, showing that ion-specific reaction dynamics fundamentally control laser-induced cavitation in aqueous environments.
\end{abstract}

\maketitle


\section{Introduction}

Laser-induced cavitation, the rapid nucleation, growth, and collapse of vapor cavities following highly localized optical energy deposition, occupies a unique intersection between fundamental fluid dynamics and a broad suite of practical technologies~\cite{young1999cavitation}. Because focused laser pulses can deposit energy at micron spatial scales and nanosecond (or shorter) temporal scales with high repeatability, laser-induced cavitation has become a preferred experimental platform for probing single-bubble dynamics, jet formation~\cite{kornfeld1944destructive, brennen2014cavitation}, shockwave emission~\cite{toker2009micro, liang2022comprehensive, horvat2022laser, pfeiffer2022thermally}, and bubble–boundary interactions~\cite{jiang2017cavitation, hironaka2023temporal, sun2023deflection, fu2023laser, fu2018experimental}. These phenomena are not merely academic: in applied settings cavitation underlies processes as diverse as minimally invasive laser surgery and ophthalmic procedures~\cite{wang2025exploratory, hellman2008biophysical, li2013single, soyama2023laser, soyama2023laser, gu2020degradation, gu2023study}, targeted drug delivery through transient cell poration~\cite{hellman2008biophysical, li2013single}, precision laser machining and micro-fabrication~\cite{soyama2023laser, zhang2024progress}, and sonochemical or photochemical enhancement of pollutant degradation in water treatment\cite{gu2020degradation, gu2023study}. In each of these application domains, the outcome of cavitation, whether it is harnessed for beneficial mechanical disruption or mitigated to protect delicate interfaces, depends sensitively on when and where cavitation occurs and on the energy available at collapse. Thus, understanding and controlling the conditions that govern cavitation initiation directly translates into improved performance, yield, and safety for materials-interface technologies that operate in real chemical environments.

A common but underappreciated feature across these application fields is that cavitation rarely occurs in idealized, ultrapure water; instead, complex chemical liquids: seawater with its multicomponent salt matrix, physiological fluids rich in buffers and biomolecules, electrolytic solutions used in electrochemical processing, and industrial process streams that contain dissolved salts and acids/bases. Dissolved ionic species and pH modifiers alter the solvent’s thermophysical~\cite{zaytsev1992properties, bialik2008boiling}, optical~\cite{dunn1991optical, aziz2020comprehensive}, and dielectric~\cite{haggis1952dielectric, hasted1948dielectric} properties in ways that are immediately relevant to optical breakdown and subsequent bubble dynamics. Changes in boiling point, vapor pressure, refractive index, absorption and scattering, dielectric response, and electron solvation kinetics can all influence (i) the propensity for the liquid to locally vaporize under a given optical fluence, (ii) the multiphoton and avalanche ionization pathways that produce seed electrons for plasma initiation, and (iii) the subsequent expansion and collapse kinetics of the nascent vapor cavity\cite{kennedy1997laser, ready2012effects, pfeiffer2022heterogeneous}. For example, laser microsurgery is often performed in or adjacent to physiological fluids whose ionic composition and pH differ substantially from laboratory water, yet the clinical efficacy and collateral tissue damage are directly affected by breakdown thresholds and bubble dynamics.
In environmental remediation and sonophotochemical processes, the presence of common anions and cations can prominently modulate the reactive species generation and thus the effectiveness of contaminant degradation.
Consequently, a predictive framework that links ionic environment, both in aggregate (ionic strength) and in ion-specific ways, to the optical breakdown threshold and to emergent cavitation behavior would deliver immediate translational value across multiple materials and interface technologies.
Despite this clear linkage, most mechanism studies of laser-induced breakdown and cavitation have emphasized the roles of laser parameters (wavelength, pulse duration, focusing geometry)~\cite{noack1998influence, vogel1999influence, tian2016stabilization, linz2015wavelength} and particulate heterogeneities~\cite{kovalchuk2010laser, li2024effect}, while systematic investigation of the roles played by dissolved ionic composition and species has been comparatively limited. 

Ionic strength provides a useful first-order descriptor of how salts influence bulk electrostatic screening and associated macroscopic solvent properties~\cite{cai2025ionic, LI_ions}. Yet ionic strength alone cannot capture the full complexity encountered in realistic fluids: decades of physical-chemical research demonstrate that ion identity (size, polarizability, hydration shell structure, and specific chemical reactivity) can exert additional, qualitatively distinct influences on interfacial and bulk processes (the so-called specific-ion or Hofmeister effects)~\cite{collins1985hofmeister, lo2012hofmeister}. In the context of laser-induced breakdown, ion-specific chemistry has the potential to alter the kinetics of short-lived intermediate species that are crucial to plasma initiation, notably hydrated electrons. These transient species form during multiphoton excitation and serve as the seeds for avalanche ionization; their formation, stabilization, and scavenging kinetics will therefore directly affect the threshold for optical breakdown.

Herein, we present a systematic experimental study that decouples non-specific electrostatic contributions, represented by ionic strength, from ion-specific chemical effects in laser-induced optical breakdown and cavitation. We design a series of controlled aqueous systems that span neutral, acidic, and alkaline regimes, enabling the isolation of purely chemical influences from collective electrostatic effects. Beyond establishing a fundamental understanding, these findings carry practical implications for applications in materials processing, sonochemistry, and biomedical laser technologies, where precise control of the ionic environment offers a powerful route to tune plasma generation and cavitation dynamics. 

\begin{figure}
\centering
\includegraphics[width = 0.97\textwidth]{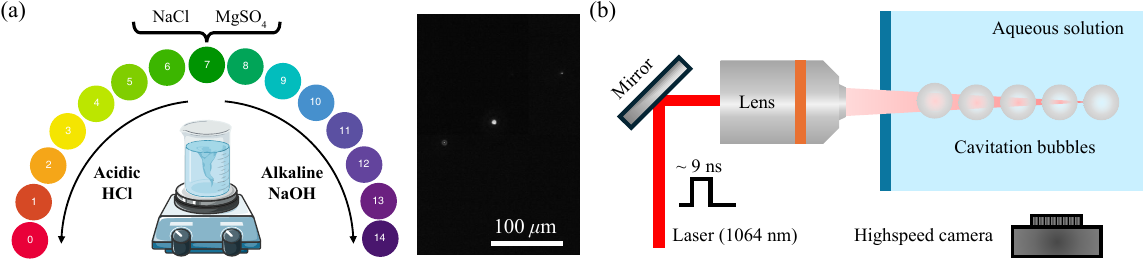}
\caption{ Experimental methods. (a) Schematic illustration of the preparation of the electrolyte solutions in this study via magnetic stirring. NTA microscopic snapshot shows trace levels of nanoscale contaminants in the formulated aqueous solution. (b) Schematic diagram of the experimental setup for pulsed laser focusing-induced optical breakdown and cavitation in electrolyte solutions. }
\label{FIG1}
\end{figure}

\section{Materials and methods}

\subsection{Materials}

High-purity DI water with a conductivity of 18.2 M$\Omega$·cm was obtained from a Milli-Q purification system (Merck, Germany) and used as the solvent for preparing aqueous solutions of various solutes. The pH (6.9) and conductivity (1.357 $\mu$S cm$^{-1}$) of this water, hereafter referred to as pure water, were measured at 25$^\circ$C. Two different salts were used in the experiment to study the effect of ionic strength under neutral solution conditions (see Figure~\ref{FIG1}(a)). Sodium chloride (NaCl, $\geq$99.999$\%$) and magnesium sulfate (MgSO$_4$, $\geq$99.99$\%$) were purchased from Thermo Scientific and Sigma-Aldrich, respectively. Hydrochloric acid (HCl, $\geq$99.999$\%$, Damas-Beta, China) and sodium hydroxide (NaOH, $\geq$99.99$\%$, Sigma Aldrich, Germany) were employed for pH adjustment after dilution to the desired concentration, and all chemicals were used without further purification. Stock solutions of NaOH, HCl, MgSO$_4$, and NaCl were prepared by gently stirring with a magnetic stirrer (HS8Pro, JoanLab, China) for 15 min at room temperature (25$^\circ$C) to ensure complete dissolution. After the solutions were prepared, a nanoparticle tracking analysis (NTA) system~\cite{ma2022measurement} was used to detect nanoscale contaminants in it to keep the nanoparticle concentration at a low level so as not to interfere with the laser-focusing path (see the snapshot in Figure~\ref{FIG1}(a)). Solution conductivity and pH were measured using a multiparameter meter (Mettler Toledo, Switzerland) before each experimental run.  

\subsection{Pulsed laser focusing-induced optical breakdown}

The experimental setup for laser-induced cavitation and visualization is illustrated in Figure~\ref{FIG1}(b). A Q-switched nanosecond-pulsed Nd:YAG laser (Dawa-100, BeamTech, China), operating at a wavelength of 1064 nm with a pulse width of 9\ ns, was employed to initiate optical breakdown in the liquid sample.
The output beam was first expanded by a telescopic beam expander to reduce divergence and improve the quality of the focal spot, and then routed through a lens assembly (beam-shaping/focusing optics) that produced a tight optical focus (focusing angle 12$^\circ$) inside a sealed, square, optical-grade glass water cell with dimensions 50 mm$\times$50 mm$\times$50 mm. The water cell was mounted on a three-axis translation stage to allow precise positioning of the focal volume relative to the cell walls.
A calibrated energy meter (FieldMaxII, Coherent, USA) was placed in the beam path and used to record the pulse energy for every shot; reported energy densities (ranging from 1.5 to $\rm{6.5\ J/cm^2}$) at the focal region were calculated from the measured pulse energy and the experimentally characterized focal spot area. Bubble nucleation, growth, and collapse dynamics were monitored optically using a high-speed camera (S1310M, Revealer, China; 150,000\ fps) synchronized with the laser trigger. The recorded field of view ($\rm{42\times3.2\ mm^2}$) was imaged at a resolution of $1152\times88$ pixels per frame. Diffuse back-illumination was employed to enhance image contrast. Three replicate measurements for each case were used to assess reproducibility and to extract statistics for threshold and dynamical parameters. Care was taken to avoid wall effects by locating the focus at the center of the pool and to maintain a constant temperature during runs. 

\section{Results and discussion}

\subsection{Cavitation nucleation in electrolyte salt solutions}

\begin{figure}
\centering
\includegraphics[width = 0.98\textwidth]{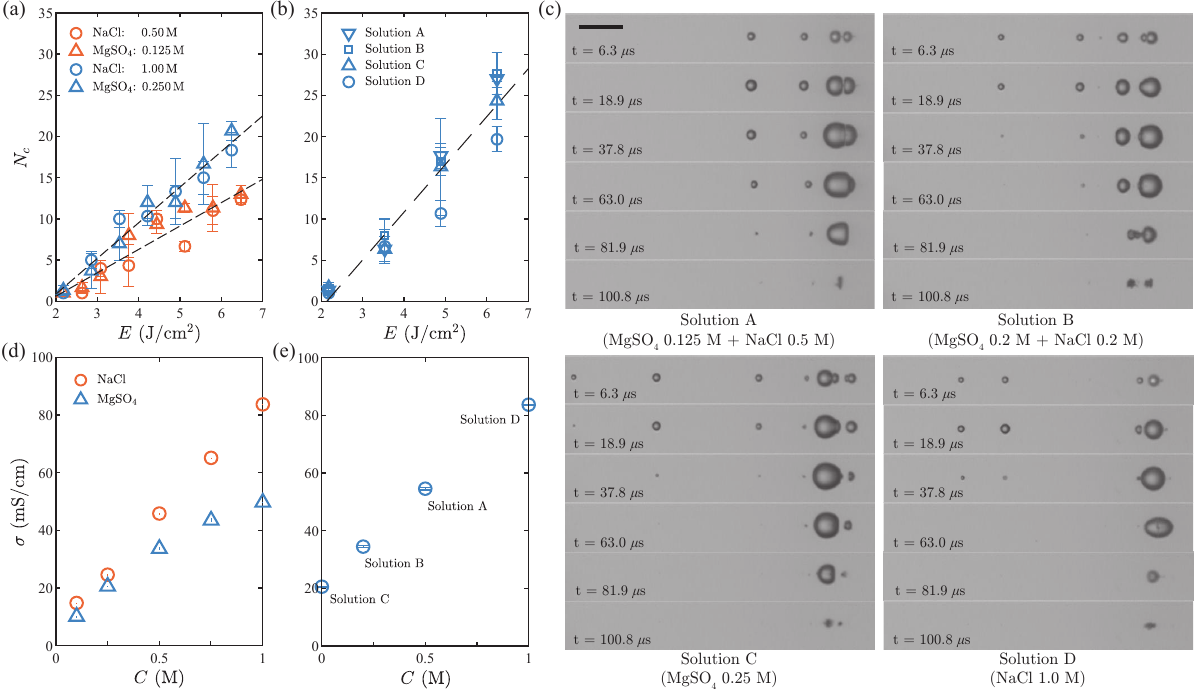}
\caption{Number of cavitation bubbles $N_{c}$ formed in (a) one-component solutions and (b) mixed solutions as a function of laser pulse energy density $E$. Error bars represent the standard error of the mean. All dashed lines represent the best linear fit. (c) Time-resolved images of laser-induced cavitation in selected solutions (corresponding to (b)) under identical ionic strength $I$. The laser pulse energy density for all cases is $\sim$6.25 J/cm$^2$. All snapshots share the same scale bar with a length of 2 mm. (d) Conductivity $\sigma$ as a function of concentration $C$ for different electrolyte solutions. (e) Comparison of conductivity in four different mixed electrolyte solutions (with NaCl concentration as the horizontal axis). }
\label{FIG2}
\end{figure}

Previous investigations into laser-induced cavitation in aqueous NaCl and $\rm MgSO_4$ solutions under tight focusing revealed two robust, reproducible trends~\cite{cai2025ionic}. First, the minimum pulse energy (or energy density) required to produce cavitation, hereafter the cavitation threshold $E_{c}$, falls rapidly as solute concentration increases, and this fall is well described by an exponential dependence on concentration (or on ionic strength). Second, the number of cavitation bubbles $N_{c}$ nucleated along the laser propagation axis increases markedly as the threshold is depressed. Although the exact positions and sizes of individual bubbles are stochastic from shot to shot, ensemble statistics show that the dominant thermodynamic variable correlating with these trends is ionic strength, defined as: 
\begin{equation}
\label{ionic}
I = \sum c_i z_i^2 / 2,
\end{equation}
where $c_i$ and $z_i$ are the molar concentration and charge number of species $i$, respectively. Figure~\ref{FIG2}(a) compares single-salt solutions of NaCl and $\rm MgSO_4$ at matched ionic strengths (0.5 M and 1.0 M). The two salts, when brought to the same $I$, generate comparable extents of cavitation: under a fixed pulse energy density, multiple bubbles form along the beam path in both cases, and the ensemble-averaged bubble count and spatial density are similar. Figure~\ref{FIG2}(b) extends this comparison to mixed NaCl/$\rm MgSO_4$ systems with different mixing ratios (Solution A: $\rm MgSO_4$ 0.125 M $\&$ NaCl 0.5 M; Solution B: $\rm MgSO_4$ 0.2 M $\&$ NaCl 0.2 M; Solution C: $\rm MgSO_4$ 0.25 M; Solution D: NaCl 1.0 M;) but identical total ionic strength $I = 1.0$ M, across a range of applied pulse energy density (2$\sim$7 J/cm$^2$). The mixed-salt data collapse onto the same cavitation-inception trend as the single-component solutions, reinforcing the conclusion that $I$ is the primary control parameter for cavitation inception in these experiments. These snapshots of the time series show that, despite differences in the solution ionic composition, these cavitation events tend to end after 100 $\mu$s (see Figure~\ref{FIG2}(c)).

Ionic strength enters directly as a convenient, concentration-weighted measure of the electrostatic influence of the dissolved species (through the $z_i$ term in Eq.~\eqref{ionic}); it compactly captures the increased ability of the solution to screen charges, to alter double-layer thicknesses at interfaces, and to modify solvation shells. These electrostatic and hydration changes can (a) destabilize nanoscopic gas pockets or pre-existing nanobubbles that act as heterogeneous nuclei, (b) change the energetics of cavity formation by modifying surface tension and local compressibility, and (c) influence the local optical/thermal response in the focal volume (for example, by modifying transient absorption or the efficiency of energy deposition into the liquid). All three effects reduce the barrier to nucleation and thereby increase the observed bubble count.

Despite the strong collapse of the data when cast in terms of ionic strength, a persistent, systematic residual was observed: NaCl solutions give slightly weaker cavitation enhancement than $\rm MgSO_4$ at the same $I$. This residual indicates measurable ion-specific contributions beyond the scalar quantity $I$. Possible sources of the difference include the higher valence and stronger hydration of $\rm Mg^{2+}$ and $\rm SO_4^{2-}$ (which affect water structure and local compressibility more strongly than $\rm Na^+$ and $\rm Cl^-$), differences in ion pairing or short-range association that alter local density fluctuations, and differential effects on surface tension or on the population of pre-existing nanoscopic cavities. In addition, multivalent ions may change the kinetics of charge screening at rapidly forming interfaces, modifying the dynamics of early-time bubble growth under the steep pressure/temperature gradients produced by a tightly focused laser pulse. In the family of inorganic salts, the effects of ion specificity are subtle or even imperceptible.

We next measured the electrical conductivity $\sigma$ of different electrolyte solutions to assess ion-specific differences in their macroscopic transport properties. Conductivity, which depends on the concentration, charge, and mobility of ions, provides an integrated measure of solute–solvent interactions. Unlike optical breakdown processes governed by transient free electrons, $\sigma$ reflects the steady-state motion of free ions under an applied field and thus characterizes equilibrium charge transport. As shown in Figure~\ref{FIG2}(d), both NaCl and $\mathrm{MgSO_4}$ solutions exhibit nearly linear increases in $\sigma$ with concentration in the dilute regime ($C < 0.5$ M), consistent with Debye–Hückel–Onsager behavior. At higher concentrations ($C > 0.5$ M), however, their trends diverge: NaCl retains higher conductivity, whereas $\mathrm{MgSO_4}$ shows a marked deviation from linearity and eventual saturation. This difference stems from ion valence and hydration effects, e.g., the strongly hydrated, divalent $\mathrm{Mg^{2+}}$ ions experience greater electrostatic coupling and reduced mobility, while the monovalent $\mathrm{Na^+}$ and $\mathrm{Cl^-}$ ions remain more mobile and dissociated. Figure~\ref{FIG2}(e) further compares solutions with identical ionic strength but different electrolytes, revealing substantial differences in $\sigma$. This confirms that ionic strength alone cannot capture ion-specific transport behavior. Conductivity instead reflects the interplay between ion charge, hydration, and correlated motion. These are macroscopic properties that define the electrolyte environment and influence, but do not directly determine laser-induced cavitation. 

\begin{figure}
\centering
\includegraphics[width = 0.50\textwidth]{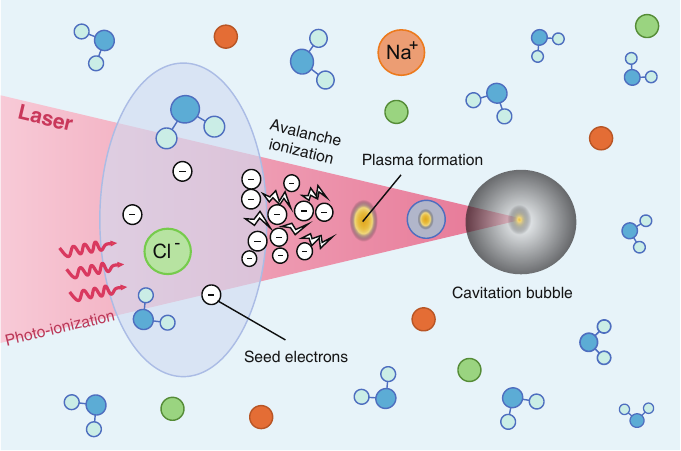}
\caption{Schematic of laser-induced plasma and cavitation bubble formation in electrolyte solution via enhanced ionization. } 
\label{FIG3} 
\end{figure}

Generally, the process of nanosecond pulsed laser-induced plasma and cavitation initiates when a high-energy laser pulse is tightly focused within the bulk, as sketched in Figure~\ref{FIG3}. The atoms or molecules at the focal point are stripped of their electrons in a very short period of time through processes such as multiphoton ionization or tunneling ionization, forming a highly localized, micrometer-scale plasma composed of free electrons and ions.
The plasma expands explosively, generating a propagating shock wave.
Concurrently, the energy deposition results in a localized low-pressure region, inducing phase change and vaporization of the water to form cavitation bubbles.
Dissolved salt ions (e.g., $\rm Cl^-$) provide abundant additional seed electrons, which remarkably enhance the subsequent avalanche ionization process, leading to efficient energy deposition.
The above results inspired us to investigate which physical micro-processions are involved in laser-induced cavitation nucleation and how their specificity affects the process.
We thus expand the electrolyte type, such as acids or bases, to study their role in cavitation inception and bubble dynamics.

\subsection{Cavitation nucleation in acidic/alkaline solutions}

\begin{figure*}
\centering
\includegraphics[width = 0.93\textwidth]{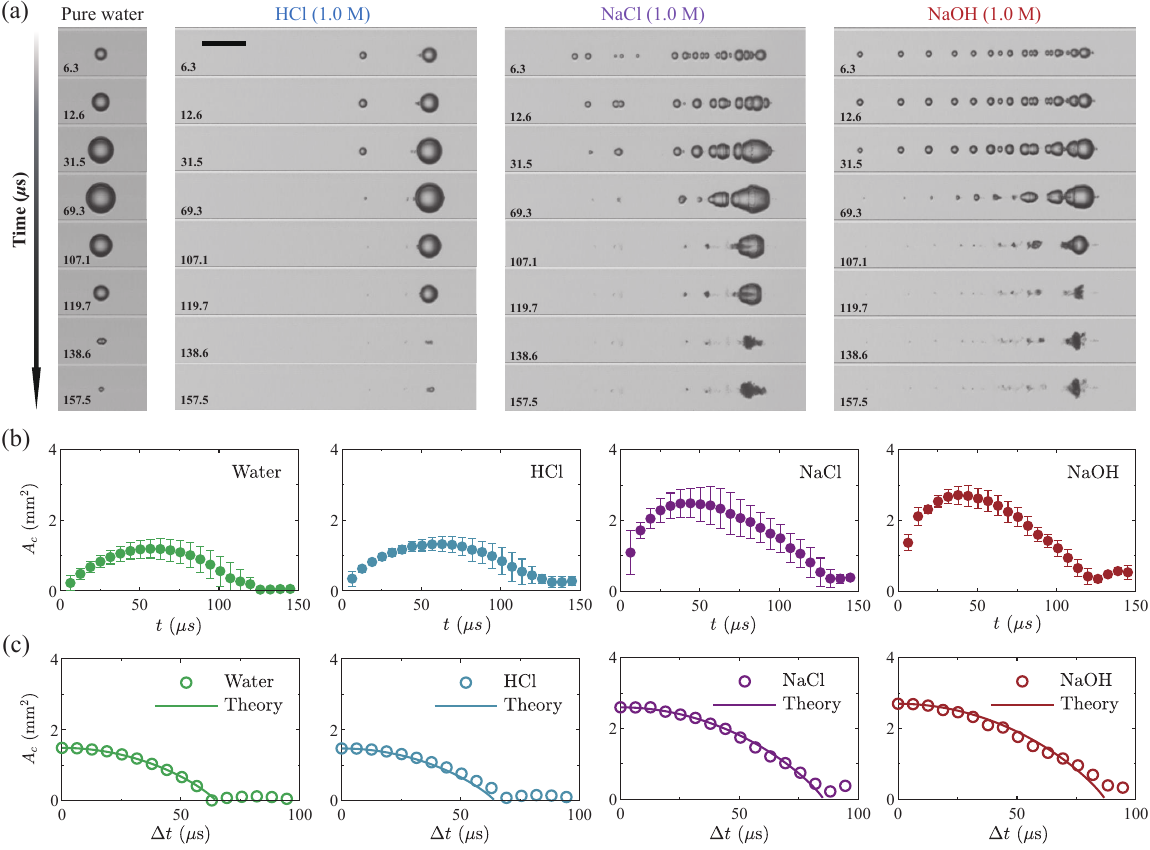}
\caption{(a) Evolution of laser-induced cavitation bubbles in ultrapure water, HCl, NaCl, and NaOH solutions.
The concentration of all electrolyte solutions was kept constant at $C = 1.0$ M. The laser pulse energy density for all cases is $\sim$6.25 $\rm{J/cm^2}$. All snapshots share the same scale bar with a length of 2\ mm. (b) Temporal evolution of the total cross-sectional area of cavitation bubbles ($A_c$) in ultrapure water and electrolyte solutions. (c) Comparison of the experimental data with the modified theoretical model for $A_c$ (solid lines). The calculation starts from the moment when the bubble features its maximum volume.} 
\label{FIG4} 
\end{figure*}

In this study, we extended our investigation from neutral salt solutions to strongly acidic (HCl) and alkaline (NaOH) solutions over a pH range of 0$\sim$14. Figure~\ref{FIG4}(a) shows cavitation at a laser energy just above the pure-water threshold: ultrapure water produces a single, nearly spherical bubble at the focus. In stark contrast, each 1.0 M electrolyte (HCl, NaCl, and NaOH) triggers multiple optical breakdown and cavitation events distributed along the beam path. This behavior matches the moving breakdown model, in which breakdown can initiate independently at multiple sites~\cite{kennedy1997laser}. The breakdown spots are spatially non-uniform: for example, cavitation is often absent downstream of the focus due to plasma shielding by upstream ionization~\cite{tian2016stabilization,fu2018experimental,fu2024secondary}. Further evidence and a detailed analysis of the shielding effect are provided in Figure S1 (see Supplemental Material).
Overall, introducing electrolytes dramatically increases the number of cavitation events, because higher ionic strength provides more seed electrons for avalanche ionization.

Even at a fixed ionic strength, the ion identity strongly influences cavitation.
At 1.0 M, the intensity follows the trend $\rm NaOH > NaCl > HCl$; notably, only two bubbles form in the HCl solution.
This ion-specific hierarchy indicates chemistry beyond mere ionic strength. The likely origin is the local pH and hydration structure.
For example, a free proton in water is small but exerts a strong attractive force on a lone pair of electrons of a water molecule, forming $\rm H_3^+$. Both $\rm H_3^+$ and $\rm OH^-$ carry tightly bound hydration shells (e.g. $\rm H_3 ^+ (H_2 O)$ and $\rm OH^-(H_2 O)_3$), which markedly distort the hydrogen-bond network. These ion-water structures affect electron dynamics: high $\rm H^+$ effectively binds electrons and raises the breakdown threshold, whereas $\rm OH^-$ supplies negative charge that boosts the free-electron population. In other words, local pH directly modulates electron ionization. Consequently, acidic solutions can inhibit optical breakdown (producing fewer bubbles) under certain conditions, while alkaline solutions enhance cavitation. This explanation is consistent with our observation that NaOH always produces the strongest cavitation and HCl the weakest, even at the same ionic strength. 

\begin{figure*}
\centering
\includegraphics[width = 0.98\textwidth]{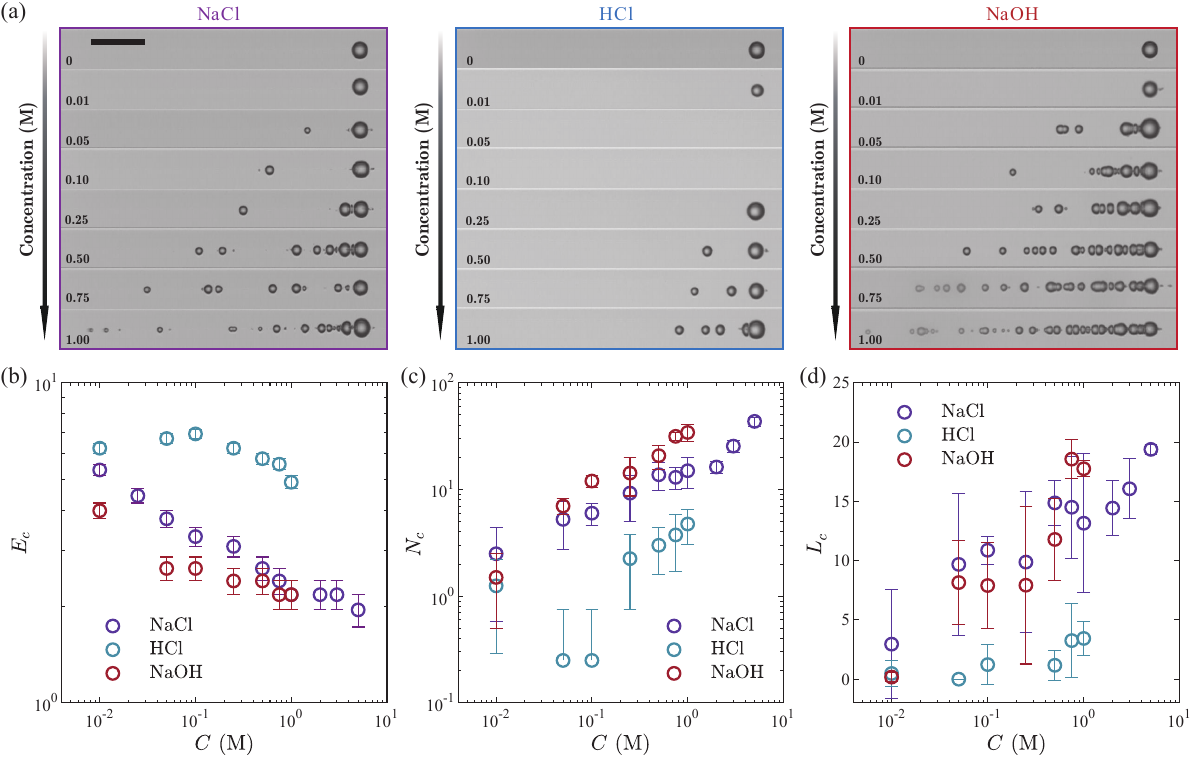}
\caption{(a) Effect of solute concentration in different electrolyte solutions (NaCl, HCl and NaOH) on cavitation bubble nucleation. These snapshots were captured 12.6 $\mu$s after the optical breakdown. All snapshots share the same scale bar of 2 mm. 
(b) Threshold of optical breakdown $E_c$, (c) cavitation bubble number $N_c$, and (d) length of cavitation zone $L_c$ along the beam path as a function of electrolyte concentration $C$ in three kinds of electrolyte solutions. Error bars represent the standard error of the mean. }
\label{FIG5}
\end{figure*}

\begin{figure*}
\centering
\includegraphics[width = 0.96\textwidth]{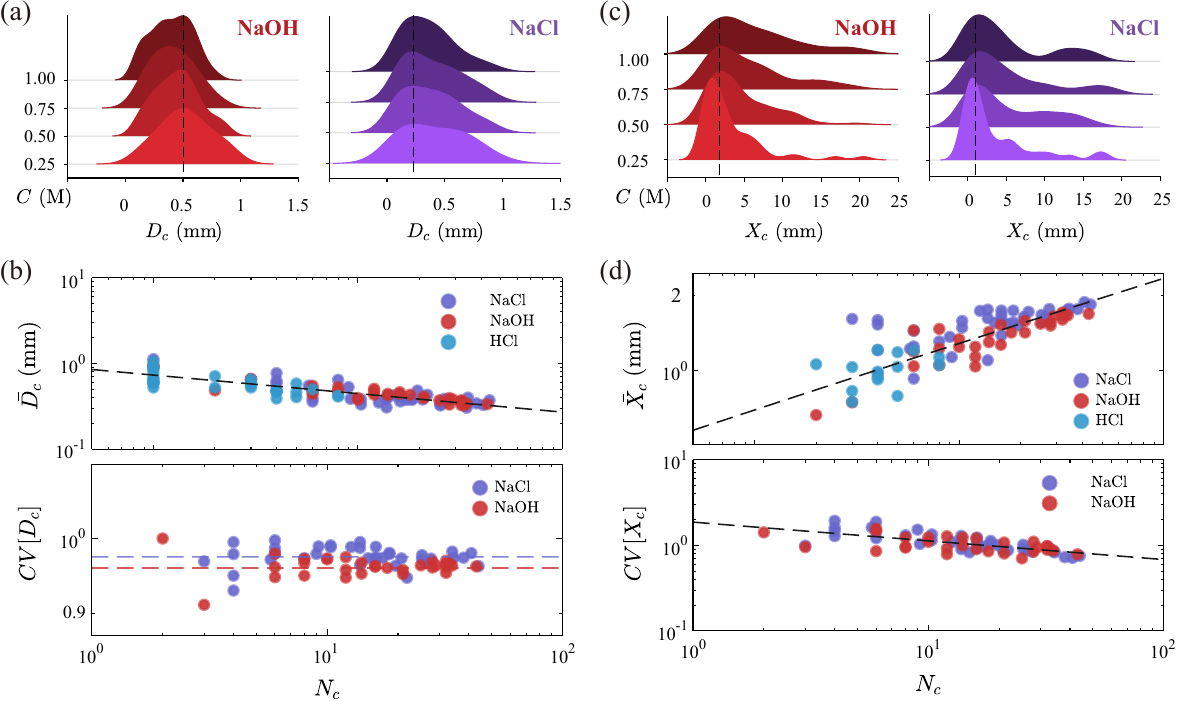}
\caption{(a) Size distribution of cavitation bubbles nucleated in solutions of NaOH and NaCl of different concentrations.  
(b) The average bubble diameter of each test (top panel) $D_c$, and the coefficient of variation (CV) of the bubble diameters (bottom panel) $CV[D_c]$, as a function of the number of cavitation bubbles, $N_c$. (c) PDF of laser-induced bubble location relative to the focal point along the beam path in NaOH and NaCl solutions. (d) The average breakdown position of each test $\bar{X}_{c}$, and coefficient of variation (CV) of bubble location for each test $CV[X_{c}]$, as a function of the number of cavitation bubbles $N_{c}$.}
\label{FIG66}
\end{figure*}

Furthermore, Figure~\ref{FIG4}(b) plots the time evolution of the total bubble cross-sectional area $A_c$ for water, HCl, NaCl, and NaOH at 1.0 M (the top panel). All cases exhibit clear growth–collapse oscillations, indicating that the fundamental cavitation dynamics (expansion followed by collapse) are preserved in every solution. The amplitude of these oscillations, however, follows the same ordering as the bubble count ($\rm NaOH > NaCl > HCl > H_2O$): NaOH produces the largest average bubble area, then NaCl, then HCl, with water the smallest.
Thus, the solution chemistry governs the intensity of each cavitation cycle while the oscillation period remains similar across cases. We also note that stronger cavitation (larger $A_c$) corresponds to more energetic collapse pulses, but the number of oscillations does not differ significantly.
Figure~\ref{FIG4}(c) compares the experimentally measured bubble dynamics with the theoretical prediction from the homobaric low-Mach model~\cite{akhatov2001collapse}. This model was specifically employed to simulate the initial contraction phase of the bubble, where the flow is characterized by a low Mach number \mbox{($|\dot{R}|/C_l < 0.3$)}. For more details of the mathematical model, see Supplemental Material (Section S2).
The computed temporal evolution of the spherical bubble radius, $R(t)$, was converted into an equivalent cross-sectional area for a direct, quantitative comparison with the experimental data via $A(t)=\pi R^2(t)$ under the axisymmetric assumption. This conversion enables a direct, quantitative comparison between the model and the experimental data, which inherently records the projected area of the bubble. The theoretical curves (dashed lines) closely match the experimental area traces for all electrolyte cases. This agreement confirms that the observed bubble growth and collapse follow the standard single-bubble physics. In other words, the differences in amplitude and number of bubbles come from the initial breakdown conditions (ion chemistry and seed-electron density), not from any change in the basic hydrodynamics of the bubble once it forms.

Figure~\ref{FIG5}(a) shows the full concentration dependence (0$\sim$6 M) of the cavitation response. The ion-specific hierarchy persists: at every ionic strength, NaOH produces stronger cavitation (more bubbles) than NaCl, and HCl remains the weakest.
For NaCl and NaOH, increasing concentration leads to generally stronger, longer cavitation (bubbles propagate farther upstream from the focus) and more breakdown sites. This is expected because adding electrolyte supplies extra charge carriers: as ionic strength rises, more seed electrons are available for avalanche ionization.
Quantitatively (Figure~\ref{FIG5}(b$\sim$d)), NaCl and NaOH follow the anticipated trends: the breakdown threshold $E_c$ falls roughly exponentially with ionic strength, while the bubble count $N_c$ and cavitation zone length $L_c$ grow sharply. These trends are consistent with the known effect that higher ionic strength amplifies laser ionization by providing additional free electrons. In particular, NaOH always shows a lower $E_c$ and higher $N_c$, $L_c$ than NaCl at the same ionic strength, indicating that $\rm OH^-$ ions are especially effective at enhancing cavitation.

By contrast, HCl exhibits a striking non-monotonic behavior. At low to moderate acid concentrations ($0.01 \sim 0.1$ M), cavitation is strongly suppressed: the threshold $E_c$ actually increases and bubble nucleation is minimal. Beyond $\sim$0.1 M, the cavitation intensity recovers toward the high-ionic-strength trend. Figure~\ref{FIG5}(b$\sim$d) shows that $E_c$ for HCl peaks around $C\sim10^{-1}$ M, whereas NaCl/NaOH simply decrease. This peak implies a competing, quenching mechanism at low $\rm [H^+]$: hydronium ions appear to actively inhibit the initiation of optical breakdown below $\sim$0.1 M. Once $\rm [H^+]$ is high enough to saturate this effect, further increases in ionic strength begin to supply electrons and $E_c$ falls. In contrast, increasing $\rm [OH^-]$ in NaOH continuously lowers $E_c$ and boosts $N_c$, $L_c$ with no such suppression. In summary, hydrogen ions can suppress cavitation-bubble nucleation under certain conditions, whereas hydroxide ions steadily promote it. These results demonstrate that ion-specific chemistry, especially local pH and hydration effects, modulates laser-induced cavitation beyond bulk ionic strength. The strong suppression of cavitation at moderate $\rm [H^+]$ and the enhanced cavitation at high $\rm [OH^-]$ cannot be explained by ionic strength alone. Instead, they reflect how protons and hydroxide alter water’s structure and free-electron chemistry. 

In Figure~\ref{FIG66}(a), we examine the size distribution of cavitation bubbles nucleated in solutions with varying electrolyte concentration. For both NaOH and NaCl solutions, the entire distribution shifts slightly toward smaller diameters at higher concentration. However, the modal size of the bubble population remains roughly constant: around 500~$\mu$m in NaOH and around 250~$\mu$m in NaCl.
This clear difference reflects anion-specific effects: hydroxide-rich (strongly alkaline) solutions tend to produce larger cavitation bubbles than chloride-rich (neutral salt) solutions under otherwise similar conditions. One possible reason is that Cl$^-$ ions, being larger and more polarizable than OH$^-$, may facilitate more efficient free-electron generation during laser breakdown. This could lead to a larger number of nucleation sites that share the available energy, resulting in individually smaller bubbles. In contrast, OH$^-$ ions have strong hydration shells that may require higher energy to perturb; the initial breakdown in NaOH may therefore generate larger primary bubbles. In addition, higher ionic strength generally increases energy dissipation (for example, through higher viscosity or enhanced thermal conductivity), which damps bubble expansion and contributes to the slight reduction in bubble size observed at high concentrations.

Figure~\ref{FIG66}(b) (top panel) shows the quantitative relationship between the mean bubble diameter $\bar D_c$ and the total number of bubbles $N_c$ across all our experiments. Remarkably, the data from NaCl, NaOH, and even HCl solutions all collapse onto a single power-law trend:
\begin{equation}
\label{eq66}
\bar D_c \propto N_c^{-1/5}.
\end{equation}
This means that as more bubbles are nucleated, the average bubble diameter systematically decreases. Physically, this scaling suggests an energy-partitioning mechanism. Each laser pulse deposits a roughly fixed amount of energy into the fluid, and that energy must be shared among the cavitation bubbles it generates. If one assumed a fixed total vaporized volume per pulse, one would expect $\bar D_c\sim N_c^{-1/3}$ (since volume scales as diameter cubed). The observed exponent of $-1/5$ is smaller in magnitude, indicating that other factors modify the simple volume-sharing picture. In particular, when many bubbles are generated in proximity, they interact strongly: the shock waves and pressure fields emitted by early-formed bubbles can compress or even shatter later ones, limiting their growth. Likewise, plasma-induced shielding means that after the first few breakdowns, the remaining laser energy available to form new bubbles is reduced.
These collective effects, shockwave interactions, and shielding, effectively diminish the energy per bubble less severely than a constant-volume model would predict, leading to a gentler $N_c^{-1/5}$ scaling instead of $N_c^{-1/3}$. 

To quantify the dispersion of the bubble diameters ($D_c$) and locations ($X_c$) within each population, we calculated their coefficient of variation ($CV$).
The $CV$ is defined as the ratio of the standard deviation to the mean, providing a normalized measure of variability.
It was calculated for each population of bubbles as: $CV[Y] = {\delta_{Y}}/{\overline{Y}} $, where $Y$ represents the variable of interest, either $D_c$ (bubble diameter) or $X_c$ (bubble location), $\delta_Y$ is the sample standard deviation, and $\overline Y$ is the sample mean.
Interestingly, despite the decrease in mean size, the coefficient of variation ($CV[D_c]$) of the bubble size distribution remains nearly constant with increasing $N_c$ (see the bottom panel of Figure~\ref{FIG66}(b)).
In other words, the relative spread of bubble diameters does not change significantly as more bubbles are produced.
This suggests that each bubble, regardless of the quantity, experiences a similar level of random growth variability.
However, we do observe a consistent offset between the two species of electrolytes: NaCl solutions exhibit a higher baseline $CV[D_c]$ (approximately $50\%$ larger) than NaOH. In practical terms, cavitation bubbles in NaCl solutions are more heterogeneous in size, whereas those in NaOH are more uniform.
This greater uniformity in the alkaline solution could arise from the more rigid and structured hydration environment around OH$^-$ ions, which may enforce a more consistent bubble nucleation process. The snapshot in Figure~\ref{FIG5}(a) confirms this: bubbles in NaOH appear more monodisperse than those in NaCl.

Next, we analyze where along the beam path the cavitation bubbles form. Figure~\ref{FIG66}(c) plots the probability distribution of bubble axial positions relative to the focal plane, for different concentrations of NaCl and NaOH. A clear trend emerges: at higher electrolyte concentration, a larger fraction of cavitation events occur farther from the focal point. In low-concentration solutions, most bubbles nucleate very near the focus where the laser intensity is highest.
As the concentration rises, however, the distribution shifts outward and even transits into a bimodal pattern (for example, in NaCl above about 0.5 M). This indicates that intense plasma formed during the first breakdowns is partially shielding the focal region.
The nascent plasma plume absorbs and scatters the laser light; thus, subsequent breakdowns occur at positions where the beam is still above threshold, which may be upstream or downstream of the original focus.
The appearance of two peaks away from the focus in concentrated solutions underscores this effect: strong plasma formation creates effectively two new “hotspots” along the beam path where cavitation preferentially occurs. Notably, although NaCl and NaOH show similar qualitative shifts, the chloride solution tends to exhibit these shielding effects at somewhat lower concentrations than NaOH. This might be because Cl$^-$-rich water produces free electrons more readily, intensifying plasma formation and thus prompting earlier onset of shielding compared to the more heavily hydrated OH$^-$ case.

Figure~\ref{FIG66}(d) (top panel) quantifies these trends by plotting the average breakdown position $\bar X_c$ versus the number of bubbles $N_c$. All electrolytes (NaCl, HCl, and NaOH) fall on a single, increasing trend: as $N_c$ grows, $\bar X_c$ moves further from the focal plane. In other words, experiments that produce many bubbles also tend to have their breakdown “center of mass” shifted downstream. This is consistent with the idea that early breakdowns near focus sap energy (via plasma formation or bubble generation), so the bulk of cavitation migrates outward when many bubbles form. The bottom panel of Figure~\ref{FIG66}(d) shows the $CV$ of the bubble locations ($CV[X_c]$) as a function of $N_c$.
We find that $CV[X_c]$ decreases approximately exponentially with $N_c$, namely $CV[X_c] \propto  N_c^{-1/5}$, meaning that as more bubbles are produced, their nucleation sites cluster more tightly around the mean position. In practical terms, this implies that large bubble clusters tend to form in a relatively localized region rather than being widely dispersed along the beam. This clustering likely occurs because once a few breakdowns establish a plasma or microbubble region, the local environment is primed for additional breakdowns (lowered threshold, gas nuclei present, etc.), drawing subsequent events into the same zone. The above results show that while the specific anion (OH$^-$ vs. Cl$^-$) influences the absolute sizes and uniformity of cavitation bubbles, the overall scaling and spatial trends are largely universal. The anion-specific differences likely stem from how OH$^-$ and Cl$^-$ ions modulate the initial plasma and bubble dynamics (through factors like electron solvation and hydration shell rigidity).

\begin{figure*}
\centering
\includegraphics[width = 0.97\textwidth]{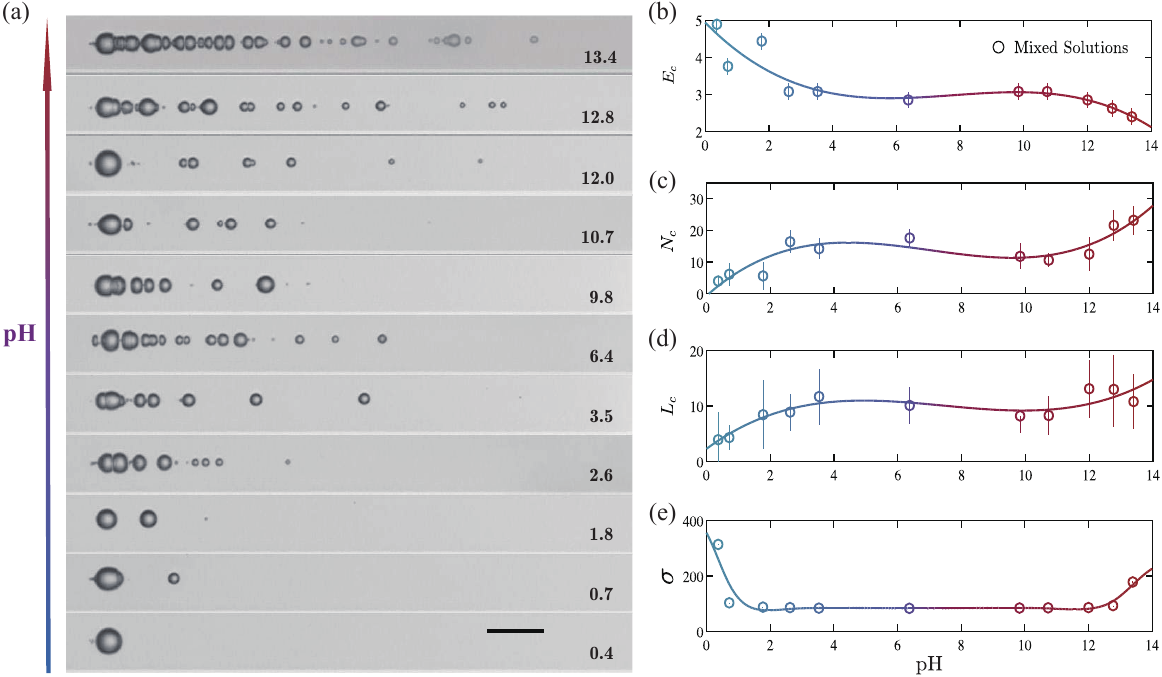}
\caption{(a) Snapshots of laser-induced cavitation bubbles in solutions at various pH levels (0$\sim$14) and constant ionic strength ($I = 1.0$ M). The scale bar presents 2\ mm. (b) Breakdown threshold of cavitation $E_{c}$, (c) number of cavitation bubbles $N_{c}$, (d) length of cavitation zone along the focus axis $L_{c}$ and (e) conductivity $\sigma$ as a function of pH levels.}
\label{FIG7}
\end{figure*}

\subsection{Decoupling the effects of ionic strength and ion specificity on cavitation nucleation}

We first note that solutions of HCl, NaCl, and NaOH at the same ionic strength produced dramatically different cavitation behavior, implying that $\rm H^+$ and $\rm OH^-$ exert specific chemical effects beyond simple charge screening.
To systematically probe this, we maintain a total ionic strength $I$ of 1.0 M while varying pH from 0 to 14 by mixing NaCl with HCl or NaOH.
For example, pH = 1 was achieved with 0.10 M HCl $\&$ 0.90 M NaCl (total [NaCl] adjusted to keep the ionic strength constant).
Figure~\ref{FIG7}(a) compares cavitation bubble nucleation along the laser focus in these solutions under fixed laser energy density. We observe that cavitation becomes more vigorous at higher pH: both the number of bubbles and the overall length of the cavitation zone increase as the pH rises.

As shown in Figure~\ref{FIG7}(b–d), the breakdown threshold $E_c$ is essentially constant from pH 3 to 12 (where [$\rm H^+$] and [$\rm OH^-$] are both below $10^{-2}$ M). However, once the solution becomes more acidic, a sharp jump in $E_c$ occurs at $\rm pH \approx 2$ (when $\rm [H^+] \geq 10^{-2}$ M). This coincides with a sudden drop in bubble counts (Figure~\ref{FIG7}(c)). Conversely, above $\rm pH \geq 12$ ($\rm [OH^-] \geq 10^{-2}$ M), $E_c$ gradually decreases and cavitation activity grows (longer $L_c$ in Figure~\ref{FIG7}(d)). Notably, the electrical conductivity remains nearly flat across most of this pH range (Figure~\ref{FIG7}(e)), indicating that bulk charge transport does not predict breakdown behavior. In fact, increasing ionic strength in general is known to lower the breakdown threshold by supplying seed electrons for avalanche ionization; by fixing the ionic strength here, we ensure that the trends we see are due to ion-specific chemistry rather than total charge density. Our data thus show that specific ions, especially $\rm H_3 O^+$ and $\rm OH^-$, are modulating cavitation under constant ionic strength.
So, how do specific types of ions participate in laser breakdown behavior?

The observed pH-dependent threshold shifts strongly suggest that ion-specific chemical kinetics, not just electrostatic screening, govern optical breakdown. Laser-induced cavitation begins with plasma formation (optical breakdown), which requires generating a critical density of free (seed) electrons. For nanosecond pulses in water, breakdown is not achieved by directly bridging the 9.5 eV bandgap~\cite{linz2015wavelength}. Rather, it proceeds via a two-step multiphoton process. In the first step, multi-photon absorption ionizes a few water molecules to produce quasi-free electrons along with $\rm H_3 O^+$ and $\rm \cdot OH$ radicals~\cite{crowell1996multiphoton, lian2004geminate}. These electrons rapidly thermalize (“hydrate”) to form solvated electrons ($\rm e_{aq}^-$). In the second step, the $\rm e_{aq}^-$ absorb additional photons (inverse Bremsstrahlung) to reach the conduction band, and then trigger an avalanche of impact ionization. Ultrafast measurements confirm this picture: breakdown initiation energies ($\sim$6.6 eV) are well below the bandgap~\cite{linz2015wavelength}, implying that excitation into a transient solvated state precedes full conduction-band ionization.

Crucially, the net yield of seed electrons depends on the competition between excitation (up-conversion) and losses (recombination and scavenging) during this process~\cite{mincher2014radiation}. In particular, any e$_{\rm aq}^-$ that avoids immediate geminate recombination can be scavenged in the bulk by excess $\rm H_3 O^+$ or $\rm \cdot OH$ via:
\begin{align}
    \begin{cases}
        \rm {e}_{aq}^{-}\ +\ \rm {H}_{3}O^{+}\ \to\  {\rm H\cdot}\ + \ \text{H}_{2}\text{O}, \\
        \rm {e}_{aq}^{-}\ +\  {\cdot \rm{OH}}\ \to\ {OH}^{-}. 
    \end{cases} 
    \label{H}
\end{align}
These reactions are both extremely fast (essentially diffusion-controlled) in water. For example, the $ \rm {e}_{aq}^{-}\ +\ \rm {H}_{3}O^{+}$ reaction has a rate constant~\cite{ma2015deciphering} of $10^{10}$–$10^{11}$ L/(mol$\cdot$s). Indeed, measurements report $k\approx4.2\times10^{10}$ Lmol$^{-1}$s$^{-1}$ (at ambient pressure) for hydronium-electron recombination. Similarly, $\rm \cdot OH$ radicals react with $\rm e_{aq}^-$ with $k\approx3.1\times10^{10}$ Lmol$^{-1}$s$^{-1}$~\cite{christensen1994temperature}. Thus $\rm H_3 O^+$ is a particularly potent quencher of e$_{\rm aq}^-$ in acidic solution, and $\rm \cdot OH$ is also a very efficient scavenger.

\begin{figure*}
\centering
\includegraphics[width = 0.96\textwidth]{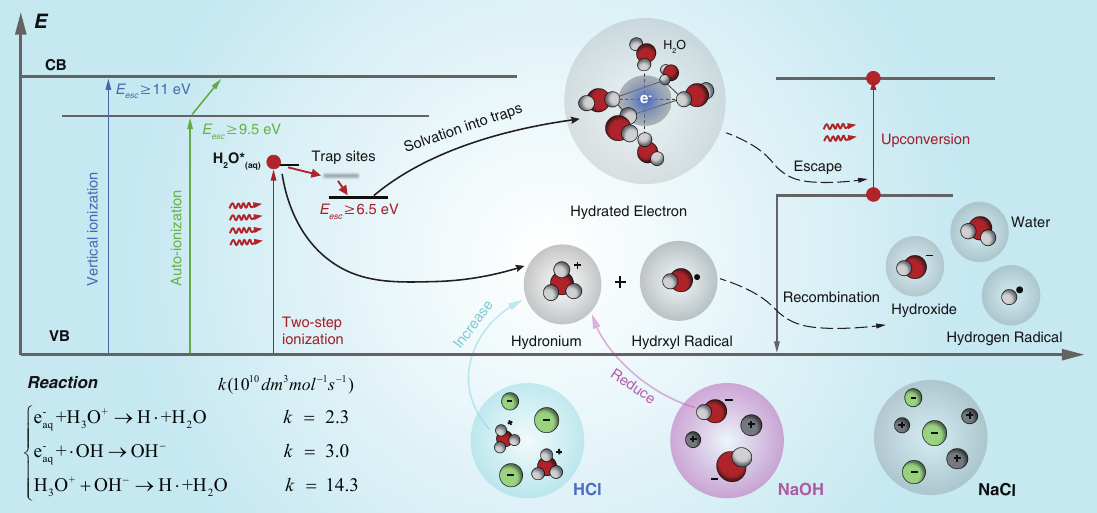}
\caption{Schematic of ionization pathways in water and ion-mediated seed electron dynamics. The schematic compares direct vertical ionization, autoionization, and the dominant two-step process for IR nanosecond pulses: multiphoton excitation ejects an electron near a pre-existing trap site, forming a solvated electron ($\rm{e_{aq}^{-}}$), a hydroxyl radical ($\rm\cdot OH$), and a hydronium ion (H$_{3}\rm{O}^{+}$). }
\label{FIG8}
\end{figure*}

In strongly acidic solutions (high [$\rm H_3 O^+$]), these scavenging pathways dominate. For instance, at pH=1 ($\rm [H_3 O^+] \approx 0.1$ M), the diffusion-limited reaction with $\rm H_3 O^+$ effectively “kills” most hydrated electrons almost immediately. This shortens the e$_{\rm aq}^-$ lifetime and severely reduces the number surviving long enough to absorb further photons. As a result, the seed-electron density available for avalanche is drastically suppressed, and a much higher laser irradiance is needed to reach the breakdown threshold. This explains the abrupt increase in $E_c$ observed once [$\rm H_3 O^+$] rises above $\sim 10^{-2}$ M (pH$\sim$2), and the concomitant drop in bubble nucleation (Fig.~\ref{FIG7}(c)). In other words, as soon as hydronium becomes a substantial fraction of the ionic species, it quenches hydrated electrons so efficiently that cavitation is inhibited.

By contrast, in highly alkaline solutions ($\rm [OH^-] \gg [H_3 O^+]$), the chemistry shifts.
Larger $\rm [OH^-]$ drives the neutralization equilibrium, 
\begin{equation}
\label{eq:OH_H_equilibrium}
\text{H}_{3}\text{O}^{+} + \text{OH}^- \rightleftharpoons  \text{H}_2\text{O},
\end{equation}
which greatly lowering the free $\rm [H_3 O^+]$.
In effect, $\rm OH^-$ removes the potent $\rm H_3 O^+ $ scavenger from the bulk (even though $\rm OH^-$ itself does not directly capture electrons). With fewer $\rm H_3 O^+ $ present, the competing quenching pathway (1) is suppressed, so more $\rm e_{aq}^-$ survive long enough to be upconverted. The net result is an increase in the seed-electron yield and hence a lower breakdown threshold.
This manifests as a gradual decrease in $E_c$ at $\rm pH > 12$, with a corresponding steady rise in bubble count. Importantly, this effect is more gradual than the acidic case, because increasing [$\rm OH^-$] reduces [$\rm H_3 O^+$] indirectly via neutralization, rather than reacting with $\rm e_{aq}^-$ itself. In summary, while a high pH does not directly provide new quenching channels, it indirectly enhances cavitation by removing hydronium. These findings confirm that laser cavitation nucleation in electrolyte solutions is not determined solely by overall ionic strength or bulk conductivity, but is critically mediated by ion-specific reaction kinetics. In particular, pH controls the balance of hydrated-electron scavenging. At low pH, the abundance of $\rm H_3 O^+$ opens a diffusion-limited quenching channel that raises the breakdown threshold sharply; at high pH, removal of $\rm H_3 O^+$ (via $\rm OH^-$) lowers the threshold gradually.
Thus, the chemical identity and reactivity of the ions, not just their charge, dictate the optical breakdown and cavitation behavior. This mechanistic picture, rooted in well-established radiolytic kinetics, explains the smooth transition we observe between the HCl and NaOH endpoints. It also highlights that conductivity (a measure of charge transfer) alone is a poor predictor of breakdown: what matters is the lifetime of seed electrons in the face of ion-specific scavenging, not just the ability of ions to carry current.

\section{Conclusions}

This study establishes a unified understanding of how electrolyte chemistry governs laser-induced optical breakdown and cavitation in water. By systematically decoupling ionic strength from ion-specific effects, we show that both electrostatic and chemical factors jointly determine cavitation inception. Increasing ionic strength consistently lowers the cavitation threshold and enhances bubble generation, owing to charge screening and the enrichment of seed electrons that facilitate avalanche ionization. Yet, this general trend alone cannot explain the pronounced differences observed across acidic and alkaline environments, revealing that ion specificity plays a decisive role in modulating breakdown behavior. 

The experiments demonstrate that $\mathrm{H^+}$ and $\mathrm{OH^-}$ ions exert opposite effects even at identical ionic strength. Acidic solutions suppress cavitation, whereas alkaline solutions enhance it. This asymmetry arises from the chemical reactivity of hydronium and hydroxide ions toward hydrated electrons ($\rm{e_{aq}^-}$), which mediate the initiation of optical breakdown. In strongly acidic media, $\mathrm{H_3O^+}$ ions rapidly scavenge $\mathrm{e_{aq}^-}$ through diffusion-controlled reactions, shortening their lifetime and reducing the seed-electron density required for avalanche growth. In contrast, $\mathrm{OH^-}$ ions indirectly extend electron survival by neutralizing $\mathrm{H_3O^+}$, thereby increasing the net electron yield and lowering the breakdown threshold. This molecular-level mechanism explains the opposing pH-dependent responses observed across the full acidity–alkalinity range.

Statistical analysis of bubble populations further reveals that, despite variations in cavitation intensity, the overall distribution characteristics remain robust across different ionic environments. The observed invariance in the relative size dispersion implies that once breakdown occurs, subsequent bubble dynamics are governed mainly by hydrodynamic interactions, such as plasma shielding and shockwave coupling, rather than by the specific ionic species present. This finding suggests a self-regulating process in which local energy competition among neighboring bubbles leads to a universal cavitation pattern.

Finally, the results confirm that bulk electrical conductivity, often regarded as an indicator of charge transport, fails to predict optical breakdown thresholds. Conductivity remains nearly constant across the wide pH range, whereas cavitation behavior varies strongly with ion chemistry. This decoupling highlights that the microscopic kinetics of hydrated-electron formation and scavenging, not macroscopic charge mobility, ultimately dictate cavitation inception. These findings bridge laser–plasma physics and aqueous chemistry, revealing that laser-induced cavitation can be rationally tuned through the ionic environment. This insight provides new strategies for controlling plasma generation and cavitation dynamics in chemically complex fluids relevant to laser processing, sonochemistry, and biomedical applications.

\begin{acknowledgments}
This work is supported by National Natural Science Foundation of China under Grants Nos. 12572290, 12202244 and 92252205, and the Oceanic Interdisciplinary Program of Shanghai Jiao Tong University (No: SL2023MS002).
\end{acknowledgments}



%

\end{document}